\documentclass[a4paper,12pt]{article}
\usepackage{amssymb,amsmath,amsfonts}
\usepackage{graphicx}
\usepackage{fullpage}
\usepackage[latin1]{inputenc}
\usepackage{url}
\usepackage{amsthm}

\title{A new lifetime model with decreasing failure rate}
\author{Wagner Barreto-Souza$^{*}$ and Hassan S. Bakouch$^{**}$
\\\\
$^{*}$Departamento de Estat\' \i stica,\\
Universidade de São Paulo,\\
Butantã, 05508-090 -- São Paulo, SP, Brazil\\
(wagnerbs@ime.usp.br)\\\\
$^{**}$Department of Mathematics,\\
Faculty of Science, Tanta University, Egypt\\
(hnbakouch@yahoo.com)
}
\date{}

\begin{document}
\maketitle

\begin{abstract}
In this paper we introduce a new lifetime distribution by compounding exponential and Poisson-Lindley distributions, named exponential Poisson-Lindley distribution. Several properties are derived, such as density, failure rate, mean lifetime, moments, order statistics and R\'enyi entropy. Furthermore, estimation by maximum likelihood and inference for large sample are discussed. The paper is motivated by two applications to real data sets and we hope that this model be able to attract wider applicability in survival and reliability. \\

{\bf keywords:} Exponential distribution; Poisson-Lindley distribution; Exponential Poisson-Lindley distribution; Order statistics; Maximum likelihood.
\end{abstract}

\section{Introduction}

Adamidis and Loukas (1998) introduced a two-parameter lifetime distribution with decreasing failure rate by compounding exponential and geometric distributions, which was named exponential geometric (EG) distribution. In the same way, Kus (2007) and Tahmasbi and Rezaei (2008) introduced the exponential Poisson (EP) and exponential logarithmic distributions, respectively.

 Recently, Chahkandi and Ganjali (2009) introduced a class of distributions, named exponential power series (EPS) distributions, by compounding exponential and power series distributions, where compounding procedure follows the same way that was previously carried out by Adamidis and Loukas (1998); this class contains the distributions mentioned before. Extensions of the EG distribution was given by Adamidis et al. (2005), Silva et al. (2010) and Barreto-Souza et al. (2010), where the last was obtained by compounding Weibull and geometric distributions. A three-parameter extension of the EP distribution was obtained by Barreto-Souza and Cribari-Neto (2009).\\

Sankaran (1970) introduced the discrete Poisson-Lindley distribution by mixing Poisson and Lindley (see Lindley, 1958) distributions. Ghitany et al. (2008) studied the zero-truncated Poisson-Lindley distribution and its applications. A random variable $N$ with this distribution has probability function given by 
\begin{eqnarray}\label{pl}
P(N=n)=\frac{\theta^{2}}{1+3\theta+\theta^{2}}\frac{2+\theta +n}{(1+\theta)^{n}},\quad n\in\mathbb{N},
\end{eqnarray}
where $\theta >0$. In this paper, we introduce a new lifetime distribution by compounding exponential and Poisson-Lindley distributions as follows: 
Let $Z_{1},Z_{2},\ldots,Z_{N}$ be a random sample following an exponential distribution with scale parameter $\beta>0$ and probability density function in the form $g(x)=\beta e^{-\beta x}$, for $x>0$, where $N$ is a discrete random variable following Poisson-Lindley distribution with probability function given by (\ref{pl}). The marginal cumulative probability function of $X=\min\{Z_{i}\}_{i=1}^{N}$ reduces to
\begin{eqnarray}\label{cdf}
F(x)=1-\frac{\theta ^{2}e^{-\beta x}}{1+3\theta +\theta ^{2}}\frac{1+\theta +\left( 2+\theta \right) \left( 1+\theta -e^{-\beta x}\right) }{%
\left( 1+\theta -e^{-\beta x}\right) ^{2}},\quad x>0,  
\end{eqnarray}
which defines the exponential Poisson-Lindley distribution. We denote a random variable $X$ with cdf given by (\ref{cdf}) as $X\sim \mbox{EPL}(\beta,\theta)$. We notice that, as Poisson Lindley distribution is not contained in class power series of distributions, our distribution is not contained in the class EPS of distributions introduced by Chahkandi and Ganjali (2009). To generate a random variable $X\sim \mbox{EPL}(\beta,\theta)$, we computate $X=F(U)^{-1}$, with $U\sim U(0,1)$, where
$$F(U)^{-1}=\log\left\{\frac{a(1+\theta)(U-1)+(b+\sqrt{\Delta_U})/2}{2+\theta+a(1-U)}\right\}^{-1/\beta},$$ 
with $\Delta_U=[b-2a(1+\theta)(U-1)]^2+4a(1+\theta)^2(U-1)[2+\theta+a(1-U)]$, $a=\theta^{-2}(1+3\theta+\theta^2)$ and $b=3+4\theta+\theta^2$.
A simple interpretation of the proposed model comes from a situation
where failure (of a device for example) occurs due to the presence
of an unknown number, say $N$, of initial defects of the same kind.
The random variables $Z$'s represent their lifetimes and each defect can be detected only after causing
failure, in which case it is repaired perfectly. Thus, the distributional assumptions
given earlier lead to the EPL distribution for modeling the time of the first failure.\\

The paper is organized as follows. In Section 2, we derive properties of the new lifetime distribution, such as density, failure rate and mean residual lifetime. Expressions for the moment generating function and moments of the EPL distribution are presented in the Section 3. Further, order statistics and its moments also are discussed in this Section. An expression for R\'enyi entropy is given in the Section 4. Estimation by maximum likelihood and inference for a random sample from EPL distribution are discussed in the Section 5, while two applications to the real data sets are presented in the Section 6. Finally, we conclude the paper in the Section 7.

\section{Density function and failure rate}

\begin{figure}[h]
	\centering
		\includegraphics[width=0.5\textwidth]{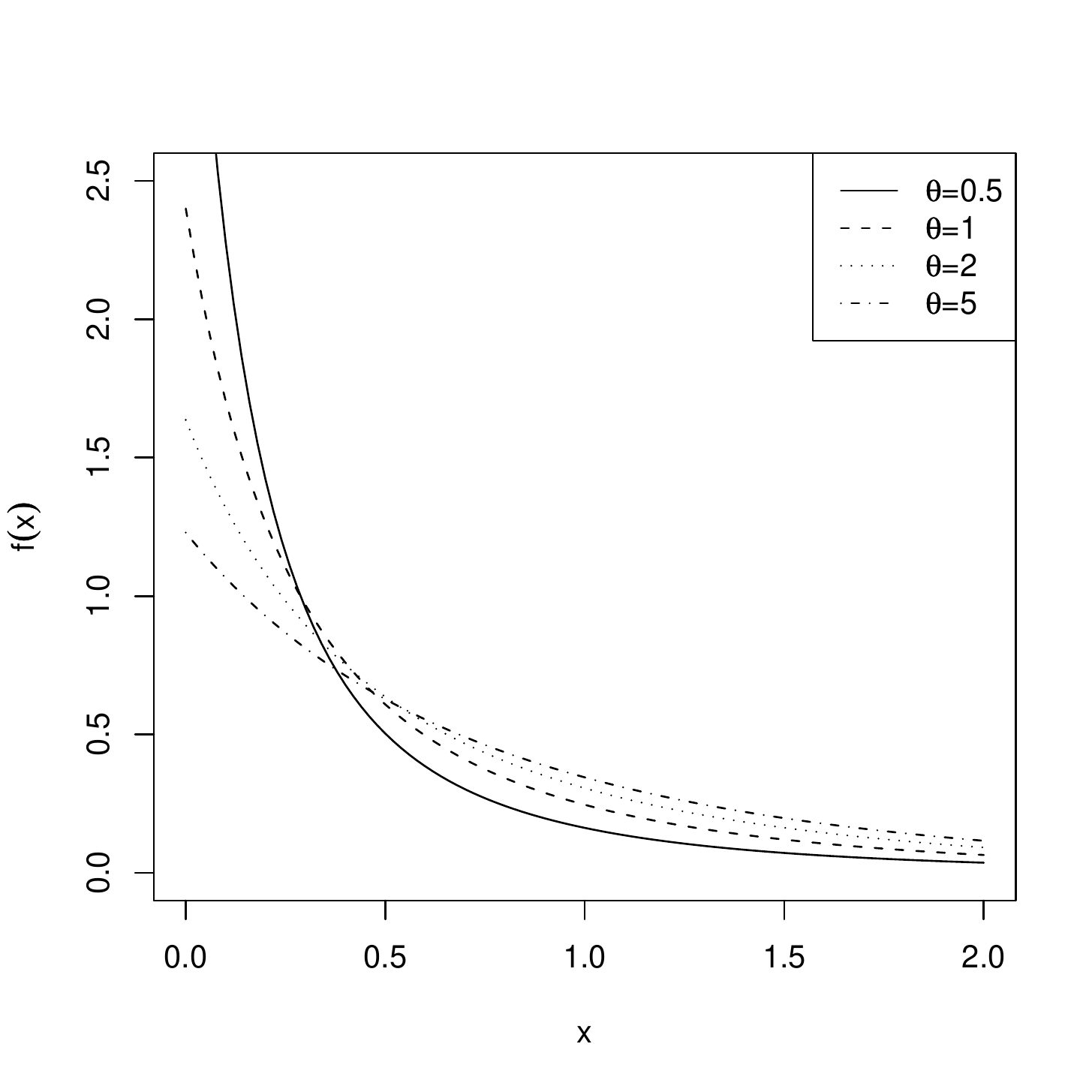}\includegraphics[width=0.5\textwidth]{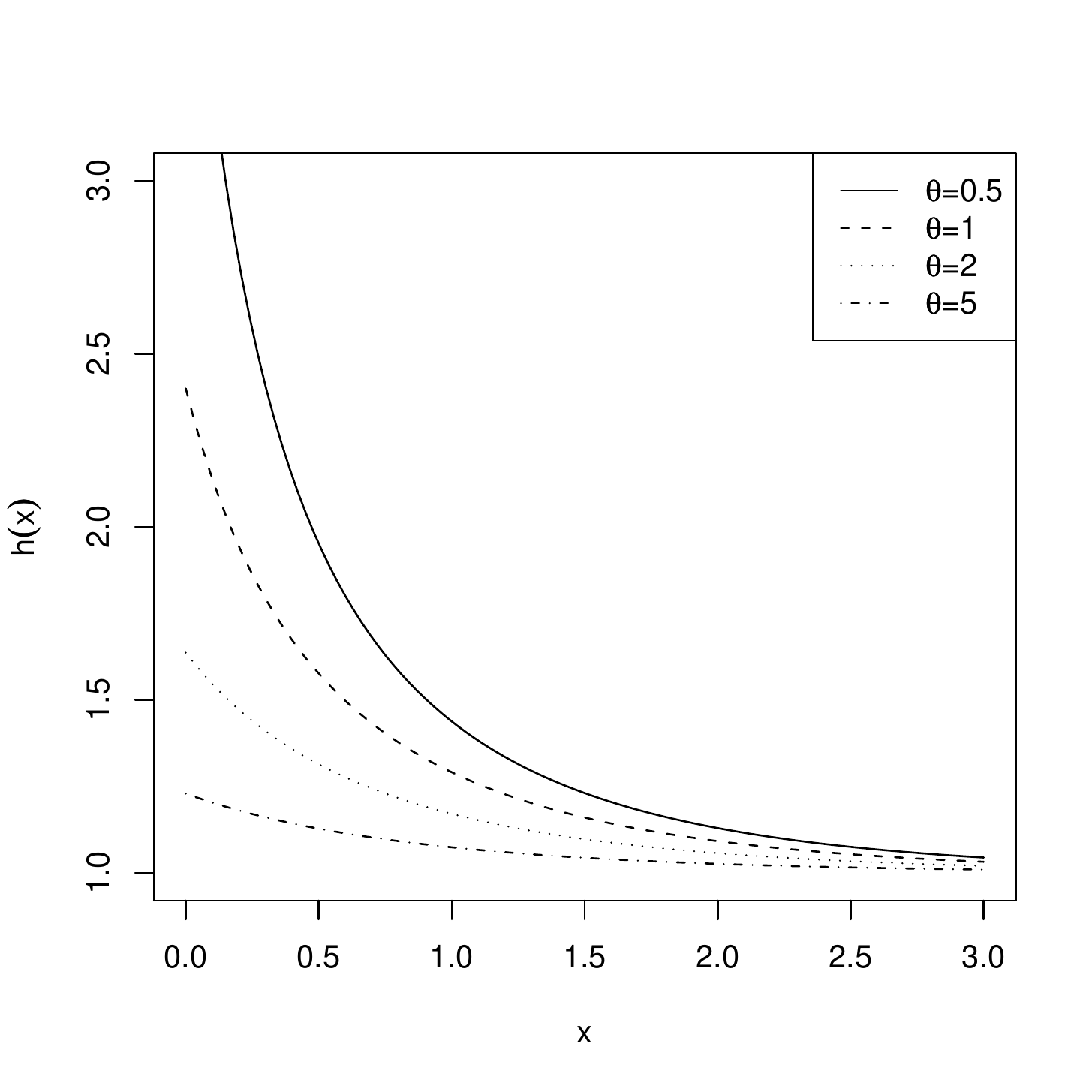}
	\caption{Plots of the pdf and failure rate of the EPL distribution for some values of the parameters.}
	\label{density}
\end{figure}

The probability density and survival functions associated to (\ref{cdf}) are given by
\begin{eqnarray}\label{pdf}
f(x)=\frac{\beta \theta ^{2}(1+\theta) ^{2}e^{-\beta x}}{%
1+3\theta +\theta ^{2}} \frac{3+\theta -e^{-\beta x}}{\left( 1+\theta
-e^{-\beta x}\right) ^{3}}
\end{eqnarray}
and
\begin{eqnarray}\label{survival}
S(x)=\frac{\theta ^{2}e^{-\beta x}}{1+3\theta +\theta ^{2}} \frac{%
1+\theta +\left( 2+\theta \right) \left( 1+\theta -e^{-\beta x}\right) }{%
\left( 1+\theta -e^{-\beta x}\right) ^{2}},
\end{eqnarray}
for $x>0$, respectively. We observe that pdf (\ref{pdf}) can be expressed as
$$f(x)=\frac{\beta\theta^2(1+\theta)}{1+3\theta+\theta^2}[g_1(x)+3g_2(x)+g_3(x)],$$
where $g_i(x)=\{(1+\theta)e^{\beta x}-1\}^{-i}$, for $x>0$ and $i=1,2,3$. Of this way, density
$f(\cdot)$ is a linear combination of monotone decreasing functions with positive coefficients and, therefore, it 
is monotone decreasing. Moreover, we have that $\lim_{x\rightarrow0^+}f(x)=\beta(\theta^3+4\theta^2+5\theta+2)/(\theta^3+3\theta^2+\theta)$ and $\lim_{x\rightarrow\infty}f(x)=0$. We also have that $f(x)\rightarrow \beta e^{-\beta x}$, when $\theta\rightarrow\infty$. With this, the exponential distribution is obtained as limiting distribution. From (\ref{pdf}) and (\ref{survival}), we obtain the failure rate of the EPL distribution:
\begin{eqnarray}\label{failure}
h(x)=\frac{\beta(1+\theta)^2(3+\theta-e^{-\beta x})(1+\theta-e^{-\beta x})^{-1}}{1+\theta+\left( 2+\theta \right)  (1+\theta -e^{-\beta x}) },\quad x>0.
\end{eqnarray}
The failure rate of the EPL distribution is decreasing according to the note of Proschan (1963) who proved that the decreasing failure rate property is
inherent to mixtures of distributions with constant failure rate. Further, we have that $\lim_{x\rightarrow0^+}h(x)=\beta(\theta^3+4\theta^2+5\theta+2)/(\theta^3+3\theta^2+\theta)$ and $\lim_{x\rightarrow\infty}h(x)=\beta$. Figure \ref{density} shows plots of the pdf (\ref{pdf}) and failure rate (\ref{failure}) with $\beta=1$ and some values of $\theta$.

Given survival to time $x_0$, the residual life is the period from time $x_0$ until the time of failure. The mean residual lifetime of a continuous distribution with survival function $S(x)$ is given by
\begin{eqnarray*}
m(x_0)=E(X-x_0|X>x_0)=\frac{1}{S(x_0)}\int_{x_0}^{\infty}S(t)dt.
\end{eqnarray*}

By taking $S(x)$ given by (\ref{survival}), we obtain the mean residual lifetime of the EPL distribution:
\begin{eqnarray*}
m(x_0)=\frac{1-(2+\theta)(1+\theta-e^{-\beta x_0})\log[1-(1-\theta)^{-1}e^{-\beta x_0}]e^{\beta x_0}}{\beta[2+\theta+(1+\theta)(1+\theta-e^{-\beta x_0})^{-1}]}.
\end{eqnarray*}
Based on the behavior of the failure function $h(x)$ of the EPL distribution, which is decreasing, we conclude that $m(x)$ is increasing (see Watson and Wells, 1961).

\section{Moments and order statistics}

As with any other distribution, many of the interesting characteristics and features of the EPL distribution can be studied through the moments.
Let $X$ be a random variable following EPL distribution with parameters $\beta>0$ and $\theta>0$. We now calculate the moment generation function of $X$ denoted by $M_X(t)=E(e^{-tX})$. For $|z|<1$ and $k>0$, we have that 
\begin{eqnarray}\label{expansion}
(1-z)^{-k}=\sum_{j=0}^\infty\frac{\Gamma(k+j)}{\Gamma(k)j!}z^j.
\end{eqnarray}
 Using this expansion in the term $\{1-(1+\theta)^{-1}e^{-\beta x}\}^{-3}$ and for $t>-\beta$, it follows that
\begin{eqnarray*}
M_X(t)&=&\frac{\beta \theta ^{2}(1+\theta)^{-1}}{
1+3\theta +\theta ^{2}}\int_0^\infty e^{-(\beta+t)x}\frac{3+\theta -e^{-\beta x}}{\{1-(1+\theta)^{-1}e^{-\beta x}\}^3}dx=\frac{\beta \theta ^{2}}{1+3\theta +\theta^{2}}\times \nonumber\\
&&\sum_{j=0}^\infty\frac{\binom{j+2}{2}}{(1+\theta)^{j+1}}
\left\{(3+\theta)\int_0^\infty e^{-[\beta(j+1)+t]x}dx-\int_0^\infty e^{-[\beta(j+2)+t]x}dx\right\}
\end{eqnarray*}
and then
\begin{eqnarray}\label{mgf}
M_X(t)=\frac{\beta\theta^2}{
1+3\theta +\theta^2}\sum_{j=0}^\infty\binom{j+2}{2}(1+\theta)^{-(j+1)}\left\{\frac{3+\theta}{\beta(j+1)+t}-\frac{1}{\beta(j+2)+t}\right\},
\end{eqnarray}
for $t>-\beta$. With formula (\ref{mgf}), we can obtain an expression for the $r$th moment by calculating $E(X^r)=(-1)^rd^rM_X(t)/dt^r|_{t=0}$. Hence, it follows that
\begin{eqnarray*}
E(X^r)=\frac{r!\,\theta^2}{
2\beta^r(1+3\theta +\theta^2)}\sum_{j=0}^\infty(1+\theta)^{-(j+1)}\left\{\frac{(3+\theta)(j+2)}{(j+1)^r}-\frac{j+1}{(j+2)^r}\right\}.
\end{eqnarray*} 
We can express the above formula in terms of the function $L_n(z)=\sum_{j=1}^\infty z^j/j^n$, which is a generalization of Euler's dilogarithm function (see, Erdelyi et al., 1953, p. 31) and is also known as the polylogarithm function; the function is quickly evaluated and readily available in standard software such as MATHEMATICA and MAPLE. So, we obtain
\begin{eqnarray}\label{moments}
E(X^r)=\frac{r!\,\theta^2}{
\beta^r(1+3\theta +\theta^2)}\left\{L_{r-1}\left(\frac{1}{\theta+1}\right)+(2+\theta)L_{r}\left(\frac{1}{\theta+1}\right)\right\}.
\end{eqnarray}
The mean and variance of $X$ are given by
\begin{eqnarray*}
\mbox{E}(X)=\frac{\theta}{\beta(1+3\theta+\theta^2)}\left\{1+\theta(2+\theta)\log\left(1+\frac{1}{\theta}\right)\right\}
\end{eqnarray*}
and
\begin{eqnarray*}
\mbox{Var}(X)=\left(\frac{\theta}{\beta}\right)^2\left\{\frac{\log\left(1+\frac{1}{\theta}\right)^2+2(2+\theta)L_2\left(\frac{1}{\theta+1}\right)}{1+3\theta+\theta^2} -\left[\frac{1+\theta(2+\theta)\log\left(1+\frac{1}{\theta}\right)}{1+3\theta+\theta^2}\right]^2\right\},
\end{eqnarray*}
respectively. Consequently, the coefficient of variation is 
\begin{eqnarray*}
\frac{\mbox{Var}(X)}{%
 \mbox{E}(X)^{2}}=\frac{2\left( 1+3\theta +\theta ^{2}\right) \left[
\log \left( 1+\frac{1}{\theta }\right) +(2+\theta )L_{2}\left( \frac{1}{%
\theta +1}\right) \right] }{\left[ 1+\theta (2+\theta )\log \left( 1+\frac{1%
}{\theta }\right) \right] ^{2}}-1. 
\end{eqnarray*} 
Table \ref{moments_num} presents mean, variance and coefficient of variation for some values of the parameters computed by using formula (\ref{moments}).
\begin{table}[!h]
\begin{center}
\begin{tabular}{c|ccc}
\hline
$\theta\downarrow$&mean&variance&cv\\
\hline
0.5& 0.4315028& 0.3923130& 2.107004\\
1& 0.6158883& 0.5966291 & 1.572898\\
5& 0.9001530& 0.8989667& 1.109458\\
10& 0.9494062&0.9491146 &  1.052966\\
\hline
\end{tabular}
\caption{
Mean, variance and coefficient of variation for $\beta=1$ and some values of $\theta$.}
\label{moments_num}
\end{center}
\end{table}
We now discuss some properties of the order statistics from EPL distribution. Order statistics are among the most fundamental tools in non-parametric statistics and inference. They enter problems of estimation and hypothesis testing in a variety of ways. Let $X_1,\ldots,X_n$ be a random sample of the EPL distribution. The pdf of the $i$th order statistics $X_{i:n}$ is given by
\begin{eqnarray*}
&&f_{i:n}\left( x\right)=\frac{n!\beta(1+\theta)^2\theta^{2(n-i+1)}e^{-\beta x(n-i+1)}(3+\theta-e^{-\beta x})}{(n-i)!(i-1)!
(1+3\theta+\theta^2)^{n-i+1}(1+\theta-e^{-\beta x})^{2(n-i)+3}}[1+\theta+\left( 2+\theta \right) \times\nonumber\\
&&\left( 1+\theta -e^{-\beta
x}\right)]^{n-i}\left[ 1-\frac{\theta ^{2}e^{-\beta x}}{1+3\theta +\theta ^{2}%
}\frac{1+\theta +\left( 2+\theta \right) \left( 1+\theta -e^{-\beta
x}\right) }{\left( 1+\theta -e^{-\beta x}\right) ^{2}}\right]^{i-1},
\end{eqnarray*}
for $x>0$. We now derive an expression for the $r$th moment of the $i$th order statistic $X_{i:n}$. For this, we use a result due to Barakat and Abdelkader (2004) and obtain
\begin{eqnarray*}
E(X_{i:n}^r)&=&r\sum_{k=n-i+1}^n (-1)^{k-n+i-1}\binom{k-1}{n-i}\binom{n}{k}\left[\frac{\theta^2(1+\theta)^{-1}}{1+3\theta+\theta^2}\right]^k\times\\
&&\int_0^\infty x^{r-1}e^{-\beta kx}\frac{\{3+\theta-(2+\theta)(1+\theta)^{-1}e^{-\beta x}\}^k}{\{1-(1+\theta)^{-1}e^{-\beta x}\}^{2k}}dx.
\end{eqnarray*}
Applying (\ref{expansion}) and binomial expansion to the terms $\{1-(1+\theta)^{-1}e^{-\beta x}\}^{-2k}$ and $\{3+\theta-(2+\theta)(1+\theta)^{-1}e^{-\beta x}\}^k$ of the above equation, we obtain an expression for the $r$th moment of $X_{i:n}$: 
\begin{eqnarray}\label{momorder}
E(X_{i:n}^r)&=&\frac{r!}{\beta^r}\sum_{k=n-i+1}^n\sum_{l=0}^k (-1)^{k-n+i+l-1}\binom{k-1}{n-i}\binom{n}{k}\binom{k}{l}
\frac{\theta^{2k}(3+\theta)^{k-l}(2+\theta)^l}{(1+\theta)^{l+k}(1+3\theta+\theta^2)^k}\times\nonumber\\
&&\sum_{j=0}^\infty\binom{j+2k-1}{2k-1} \frac{(1+\theta)^{-j}}{(k+l+j)^r}.
\end{eqnarray}
Table \ref{t1} lists some numerical values for the first four moments
of order statistics $X_{1:20}$, $X_{10:20}$ and $X_{20:20}$ from equation
(\ref{momorder}), with the index $j$ stopping at $100$, and using numerical integration. The parameter values are taking as
$\beta=0.1$ and $\theta=0.5$. The results in this table show good agreement between the two methods.
\begin{table}[!h]
\begin{center}
\begin{tabular}{cc|cccc}
\hline
$X_{i:20}\downarrow$&$r$th moment$\rightarrow$&$r=1$&$r=2$&$r=3$&$r=4$\\
\hline
$i=1$&Expression (\ref{momorder})&0.1172126 & 0.03121316&  0.01280193 &0.007203924 \\
&Numerical&  0.1260696&0.03281421 &  0.01323651& 0.007361332\\
\hline
$i=10$&Expression (\ref{momorder})&2.279001 & 4.940849& 13.43999 &  42.12036\\
&Numerical& 2.056653 & 4.899393 & 13.42872& 42.11641\\
\hline
$i=20$&Expression (\ref{momorder})& 22.10733&621.2616 & 21864.98& 944798.2\\
&Numerical&  22.10734 &621.2616 &  21864.98&944798.2 \\
\hline
\end{tabular}\caption{First four moments of some order statistics from (\ref{momorder})
and via numerical integration.}\label{t1}
\end{center}
\end{table}
We now are interested in the asymptotic distributions of the extreme values $X_{1:n}$ and $X_{n:n}$. 
 From equations (\ref{cdf}) and (\ref{survival}), it can be seen using
L'H\^opital's rule that 
\begin{eqnarray*}
\lim_{t\rightarrow 0^+}\frac{F(tx)}{F(t)}=x \quad\mbox{and}\quad \lim_{t\rightarrow \infty }\frac{S(t+x)}{S(t)}=e^{-\beta x}.
\end{eqnarray*}
From this and making use of Theorem 1.6.2 of Leadbetter et al. (1987), we
obtain
\begin{eqnarray*}
P\left(a_{n}(X_{n:n}-b_{n})\leq x\right) =e^{-e^{-\beta x}}\quad\mbox{and}\quad
P\left(c_{n}(X_{1:n}-d_{n})\leq x\right) =1-e^{-x},
\end{eqnarray*}
as $n\rightarrow \infty ,$ where the norming constants $a_{n},b_{n},c_{n}>0$
and $d_{n}$ can be determined by Corollary 1.6.3 in Leadbetter et al.
(1987). For instance, we can select $a_{n}=1$ and $b_{n}=F(1-1/n)^{-1}$ as $n\rightarrow\infty$.

\section{R\'enyi entropy}

The entropy of a random variable $X$ is a measure of uncertainty 
variation. The R\'enyi entropy is defined as $I_R(\alpha)=(1-\alpha)^{-1}\log(\int_{\mathbb{R}} f^\alpha(x)dx)$, $\alpha>0$ and $\alpha\neq1$. 
Let $f(\cdot)$ be the pdf of the $\mbox{EPL}(\beta,\theta)$ distribution. Using expansion (\ref{expansion}) for $(1+\theta-e^{-\beta x})^{3\alpha}$ and after some algebra, we obtain that
\begin{eqnarray*}
\int_0^\infty f^\alpha(x)dx=\frac{\beta^{\alpha-1}[\theta(1+\theta)]^{2\alpha}}{\Gamma(3\alpha)(1+3\theta+\theta^2)^\alpha}\sum_{j=0}^\infty&&\frac{\Gamma(3\alpha+j)(3+\theta)^{2\alpha+j}}{(1+\theta)^{3\alpha+j}j!}\times\\
&&\left[1-B_{\frac{2+\theta}{3+\theta}}(\alpha,\alpha+j-1)\right],
\end{eqnarray*}
where $B_x(a,b)=\int_0^x t^a(1-t)^{b-1}dt$ is the incomplete beta function, for $a,b>0$ and $x\in(0,1)$. Hence, an expression for R\'enyi entropy of the EPL distribution is given by
\begin{eqnarray*}
I_R(\alpha)&=&-\log\beta+(1-\alpha)^{-1}\left\{{2\alpha}\log[\theta(1+\theta)]-\log\Gamma(3\alpha)-\alpha\log(1+3\theta+\theta^2)\right\}+\\
&&(1-\alpha)^{-1}\log\left\{\sum_{j=0}^\infty\frac{\Gamma(3\alpha+j)(3+\theta)^{2\alpha+j}}{(1+\theta)^{3\alpha+j}j!}\left[1-B_{\frac{2+\theta}{3+\theta}}(\alpha,\alpha+j-1)\right]\right\}.
\end{eqnarray*}

It is important to derive an expression for the R\'enyi entropy because
it serves as a measure of the shape of a distribution and can be used to
compare the tails and shapes of various frequently used densities; for more details,
see Song (2001).

\section{Estimation and inference}

Let $X_1,\ldots,X_n$ be a random sample from $\mbox{EPL}(\beta,\theta)$ distribution with observed values $x_1,\ldots,x_n$. The log-likelihood function is given by
\begin{eqnarray}\label{loglik}
\ell_n\equiv\ell_n(\beta,\theta)&=&n\log \beta +2n\log \theta
(1+\theta )-n\log \left( 1+3\theta +\theta ^{2}\right) -\beta
\sum_{i=1}^{n}x_{i}+  \notag \\
&&\sum_{i=1}^{n}\log \left( 3+\theta -e^{-\beta x_{i}}\right)
-3\sum_{i=1}^{n}\log \left( 1+\theta -e^{-\beta x_{i}}\right) .
\end{eqnarray}

The score function associated to log-likelihood (\ref{loglik}) is $U_n\equiv U_n(\beta,\theta)=(\partial \ell_n/\partial\beta,\partial \ell_n/\partial\theta)^\top$, where
\begin{equation*}
\frac{\partial\ell_n}{\partial \beta }=\frac{n}{\beta }%
-\sum_{i=1}^{n}x_{i}+\sum_{i=1}^{n}x_{i}e^{-\beta x_{i}}\left[ \frac{1}{%
3+\theta -e^{-\beta x_{i}}}-\frac{3}{1+\theta -e^{-\beta x_{i}}}\right]
\end{equation*}
and
\begin{equation*}
\frac{\partial\ell_n}{\partial \theta }=\frac{2n\left( 1+2\theta \right) }{%
\theta (1+\theta )}-\frac{n\left( 3+2\theta \right) }{1+3\theta +\theta ^{2}}%
+\sum_{i=1}^{n}\left[ \frac{1}{3+\theta -e^{-\beta x_{i}}}-\frac{3}{1+\theta
-e^{-\beta x_{i}}}\right].
\end{equation*}

The maximum likelihood estimators of $\beta$ and $\theta$ are obtained by solving numerically the nonlinear
system of equations $U_n=0$. It is usually more convenient, however, to use a nonlinear optimization algorithm (such as the quasi-Newton algorithm known as BFGS) to numerically maximize the log-likelihood function in (\ref{loglik}).  Moreover, under regularity conditions, the expected value of the score function vanishes. With this, we obtain
\begin{eqnarray*}
E\left(\frac{1}{3+\theta -e^{-\beta X}}-\frac{3}{1+\theta -e^{-\beta X}}%
\right) =\frac{3+2\theta }{1+3\theta +\theta ^{2}}-\frac{2+4\theta }{\theta
+\theta ^{2}}
\end{eqnarray*}
and
\begin{eqnarray*}
E\left[ Xe^{-\beta X}\left( \frac{1}{3+\theta -e^{-\beta X}}-\frac{3}{%
1+\theta -e^{-\beta X}}\right) \right] =\frac{\theta[1+\theta (2+\theta )\log \left( 1+\frac{1}{\theta }%
\right)]}{\beta(1+3\theta+\theta ^{2})}-\frac{1}{\beta}.
\end{eqnarray*}%

For inference, Fisher's information matrix is required. It is given by
\begin{align*}\label{fisher}
K_n \equiv K_n(\beta,\theta) = n\left[ \begin{array}{cc}
\kappa_{\beta,\beta}&\kappa_{\beta,\theta}\\
\kappa_{\beta,\theta}&\kappa_{\theta,\theta}\\
\end{array}\right],
\end{align*}
where 
\begin{eqnarray*}
&&\kappa_{\beta ,\beta }=\frac{1}{\beta ^{2}}+\left( 3+\theta \right) E\left[
\frac{X^2e^{-\beta X}}{(3+\theta -e^{-\beta X})^2}\right]
-3\left( 1+\theta \right) E\left[ \frac{X^2e^{-\beta X}}{(1+\theta
-e^{-\beta X})^2}\right],\\
&&\kappa_{\theta,\theta }=\frac{2\left( 1+2\theta +2\theta ^{2}\right) }{\theta
^{2}(1+\theta )^{2}}-\frac{7+6\theta +2\theta ^{2}}{\left( 1+3\theta +\theta
^{2}\right) ^{2}}+E\left[ \left( \frac{1}{3+\theta -e^{-\beta X}}\right) ^{2}%
\right]-3\times\\
&&E\left[ \left( \frac{1}{1+\theta -e^{-\beta X}}\right)^{2}\right],\,\,
\kappa_{\theta ,\beta }=E\left[ \frac{Xe^{-\beta X}}{(3+\theta -e^{-\beta
X})^2}\right] -3E\left[\frac{Xe^{-\beta X}}{(1+\theta
-e^{-\beta X})^2}\right].
\end{eqnarray*}
The above expectations can be obtained numerically. Under conditions that are fulfilled for parameters in the interior of the parameter space but not on
the boun\-dary, we have that
\begin{equation*}
\sqrt{n}\left( 
\begin{array}{c}
\hat{\theta}-\theta \\ 
\hat{\beta}-\beta%
\end{array}%
\right) \overset{d}{\rightarrow }\mathcal{N}\left( \left( 
\begin{array}{c}
0 \\ 
0%
\end{array}%
\right) ,K\left(\beta,\theta \right)^{-1} \right),
\end{equation*}
as $n\rightarrow\infty$, where $K(\beta,\theta)=\lim_{n\rightarrow\infty}n^{-1}K_n(\beta,\theta)$.
The bivariate normal distribution with mean $(0,0)^\top$ and covariance matrix $K_n(\beta,\theta)^{-1}$
can be used to construct confidence intervals for the model parameters.

\section{Applications}

In this Section we fit exponential Poisson-Lindley distribution to two real data sets and compare with the exponential-geometric, exponential Poisson, exponential logarithmic, Weibull and gamma distributions, whose densities are given by
\begin{eqnarray*}
&&f_{eg}(x;\beta_1,p_1)=\beta_1(1-p_1)e^{-\beta_1 x}(1-p_1e^{-\beta_1x})^{-1}, \quad \beta_1>0, \quad p_1\in(0,1),\\
&&f_{ep}(x;\beta_2,\lambda)=\frac{\lambda\beta_2}{1-e^{-\lambda}}e^{-\lambda-\beta_2x+\lambda\exp(-\beta_2x)},\quad \beta_2,\lambda>0,\\
&&f_{el}(x;\beta_3,p_2)=\frac{1}{-\log p_2}\frac{\beta_3(1-p_2)e^{-\beta_3x}}{1-(1-p_2)e^{-\beta_3x}}, \quad \beta_3>0, \quad p_2\in(0,1),\\
&&f_{w}(x;\beta_4,\alpha)=\alpha\beta_4^\alpha x^{\alpha-1}e^{-\beta_4 x}, \quad \beta_4,\alpha>0,\\
&&f_{g}(x;\beta_5,\gamma)=\frac{\beta_5^\gamma}{\Gamma(\gamma)}x^{\gamma-1}e^{-\beta_5 x}, \quad \beta_5,\gamma>0,
\end{eqnarray*}
for $x>0$, respectively.
The first data set is given by Linhart and Zucchini (1986), which represents the
failure times of the air conditioning system of an airplane: 23, 261, 87, 7, 120, 14, 62, 47, 225, 71, 246, 21, 42, 20, 5, 12, 120, 11, 3, 14, 71, 11, 14, 11, 16, 90, 1, 16, 52, 95.

In the second data set, we consider vinyl chloride data obtained from clean upgradient monitoring wells in mg/L; this data set was used for Bhaumik et al. (2009). The data are: 5.1, 1.2, 1.3, 0.6, 0.5, 2.4,  0.5, 1.1, 8.0, 0.8, 0.4, 0.6, 0.9, 0.4, 2.0, 0.5, 5.3, 3.2, 2.7, 2.9, 2.5, 2.3, 1.0, 0.2, 0.1, 0.1, 1.8, 0.9, 2.0, 4.0, 6.8, 1.2, 0.4, 0.2.

Tables \ref{tab1app} and \ref{tab2app} give us the fitted parameters, Kolmogorov-Smirnov (K-S) statistics and its respective p-values for the first and second data sets, respectively. We see that all the distributions in both tables show a good fit for the given data sets and, in each case, we do not reject the hypothesis that the data comes from distribution considered at any usual significance level. We also see that EPL distribution proves to be a good competitor for other distributions already known in literature. 
\begin{table}[!h]
\begin{center}
\begin{tabular}{cccc}
\hline
Distribution& Estimates & K-S statistic & p-value \\
\hline
EPL$(\beta,\theta)$&(0.0101, 0.9193)&  0.1290 &0.6531\\
EG$(\beta_1,p_1)$& (0.0102, 0.6148)&0.1262 & 0.6793\\
EP$(\beta_2,\lambda)$ & (0.0106, 1.7941)& 0.1472& 0.4890\\
EL$(\beta_3,p_2)$& (0.0111, 0.1932)& 0.1288 & 0.6555\\
W$(\beta_4,\alpha)$&(0.0183, 0.8533)& 0.1531& 0.4394\\
G$(\beta_5,\gamma)$&(0.0136, 0.8135)& 0.1694 &0.3187\\
\hline
\end{tabular}
\caption{Estimates of the parameters, Kolmogorov-Smirnov statistics and its respectively p-values for the first data set.}
\label{tab1app}
\end{center}
\end{table}

\begin{table}[!h]
\begin{center}
\begin{tabular}{cccc}
\hline
Distribution& Estimates & K-S statistic & p-value \\
\hline
EPL$(\beta,\theta)$&(0.4796, 5.0811)& 0.0882 &0.9331\\
EG$(\beta_1,p_1)$& (0.4818, 0.1771)& 0.0876 & 0.9360\\
EP$(\beta_2,\lambda)$ & (0.4767, 0.4276)& 0.0880 &0.9341\\
EL$(\beta_3,p_2)$& (0.4867, 0.7022)& 0.0870 & 0.9394\\
W$(\beta_4,\alpha)$&(0.5296, 1.0101)&  0.0918 &0.9116\\
G$(\beta_5,\gamma)$&(0.5654, 1.0626)& 0.0973 & 0.8733\\
\hline
\end{tabular}
\caption{Estimates of the parameters, Kolmogorov-Smirnov statistics and its respectively p-values for the second data set.}
\label{tab2app}
\end{center}
\end{table}

\begin{figure}[h!]
	\centering
		\includegraphics[width=0.5\textwidth]{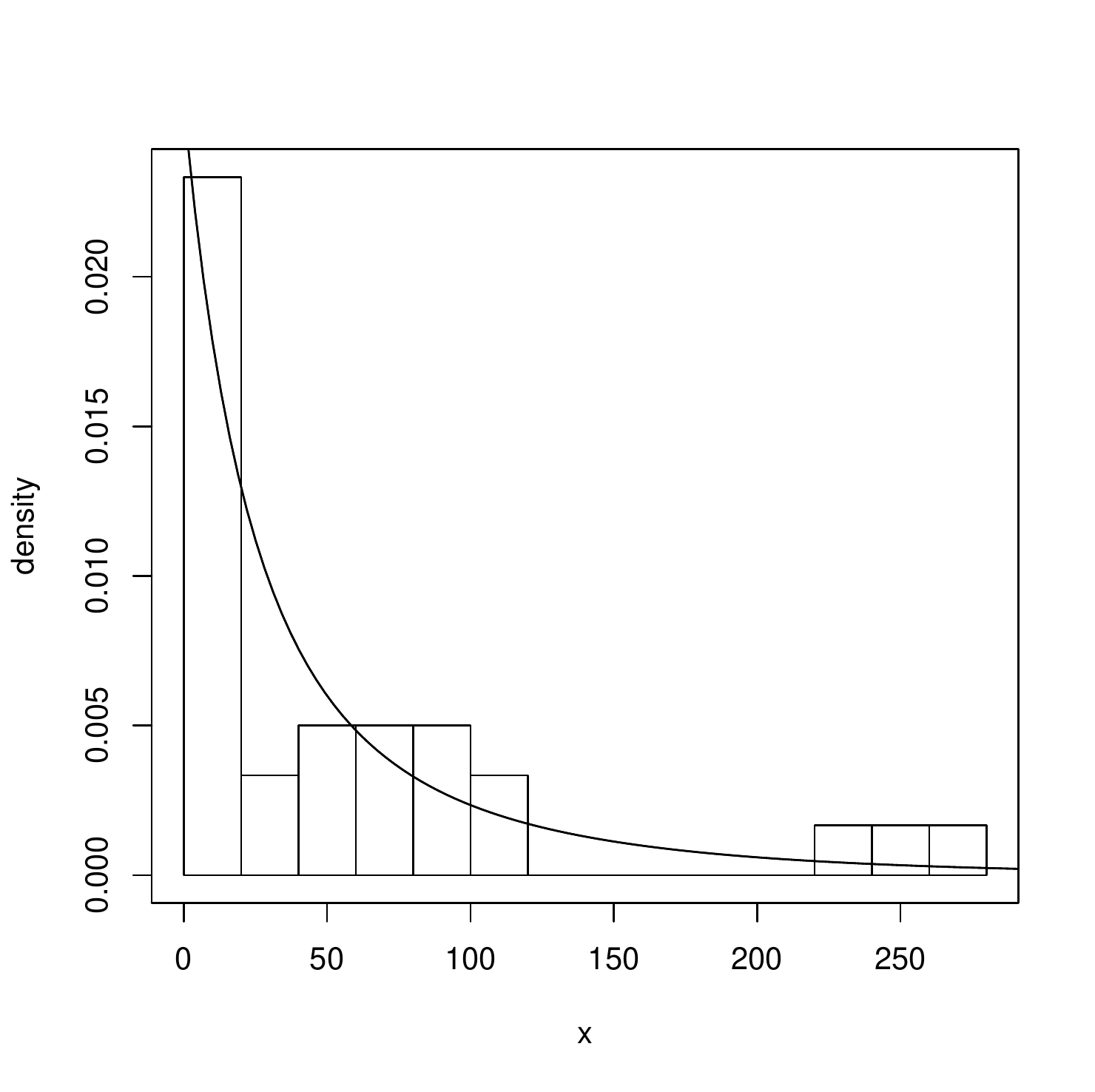}\includegraphics[width=0.5\textwidth]{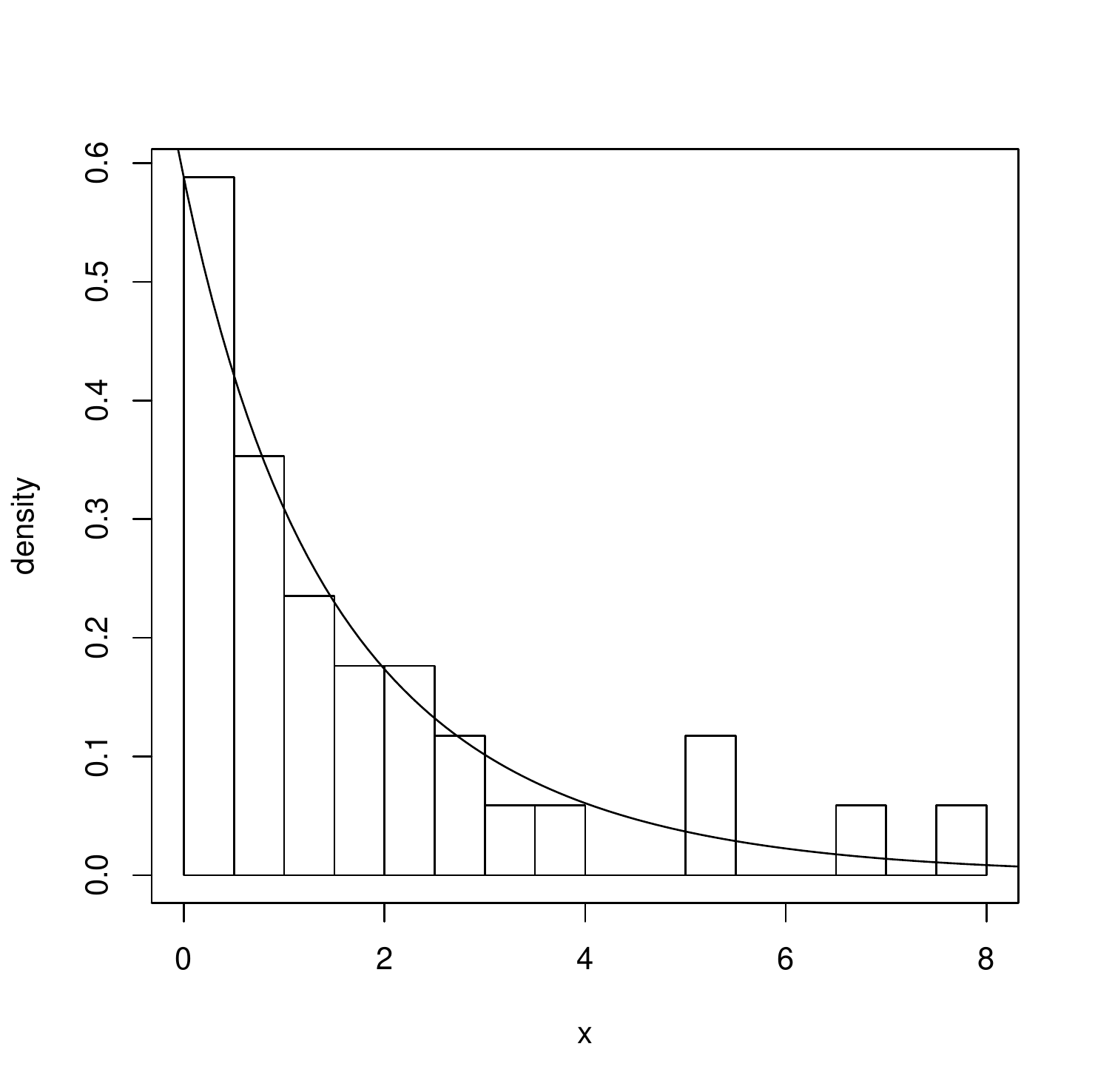}
	\caption{Plots of the fitted densities of the EPL distributions for the first (left) and second (right) data sets.}
	\label{applfig1}
\end{figure}

\begin{figure}[h!]
	\centering
		\includegraphics[width=0.5\textwidth]{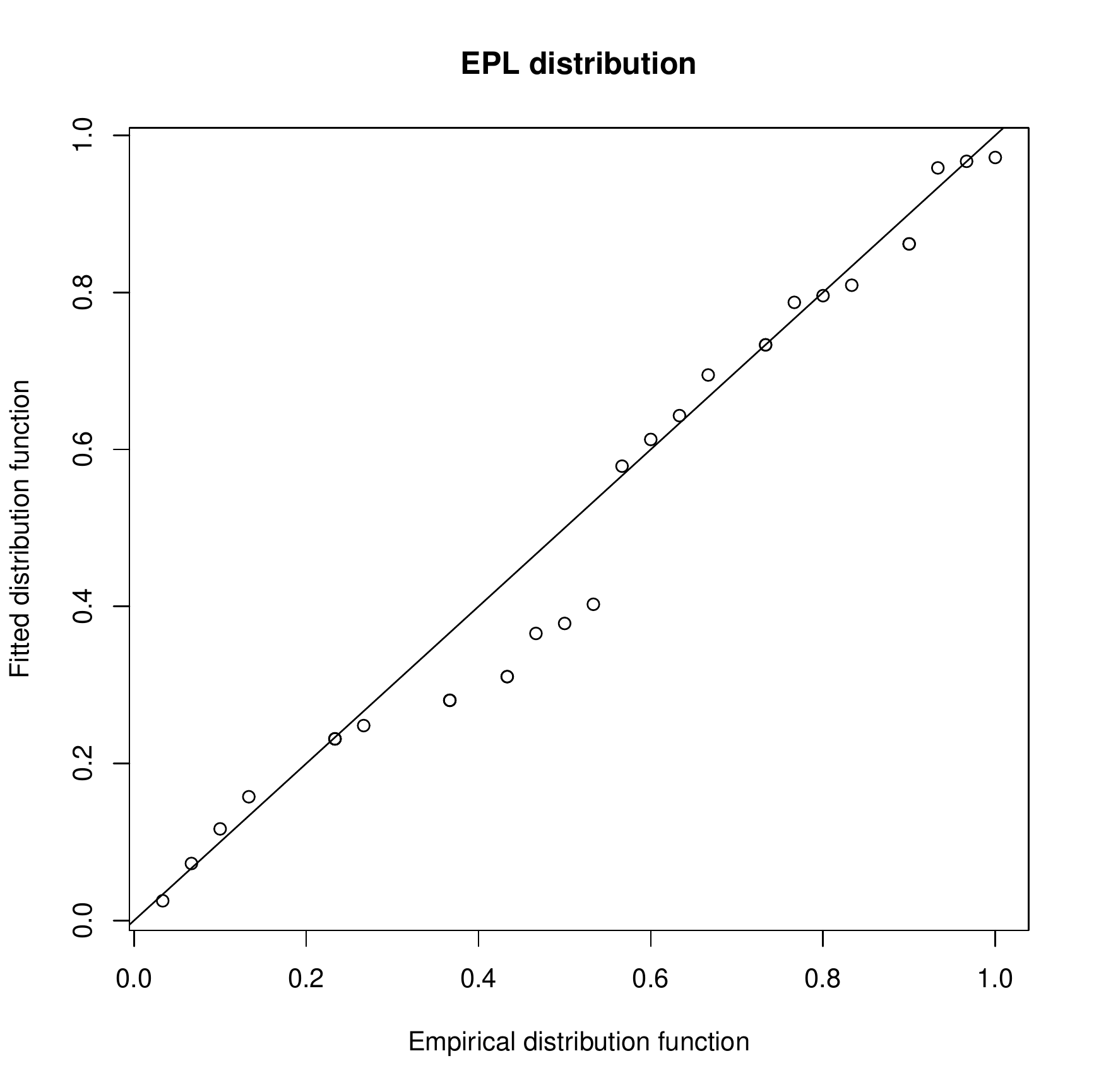}\includegraphics[width=0.5\textwidth]{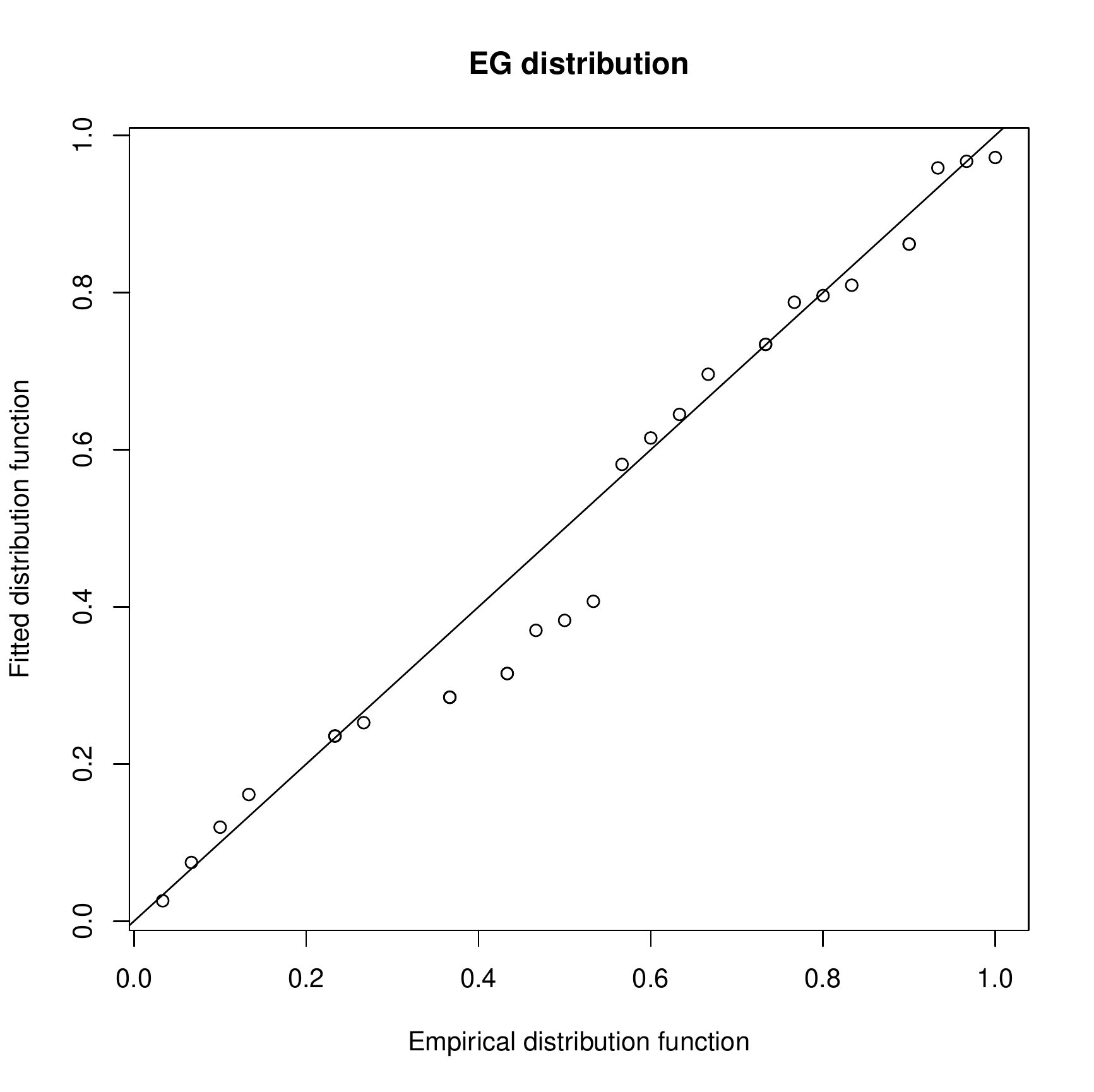}
		\includegraphics[width=0.5\textwidth]{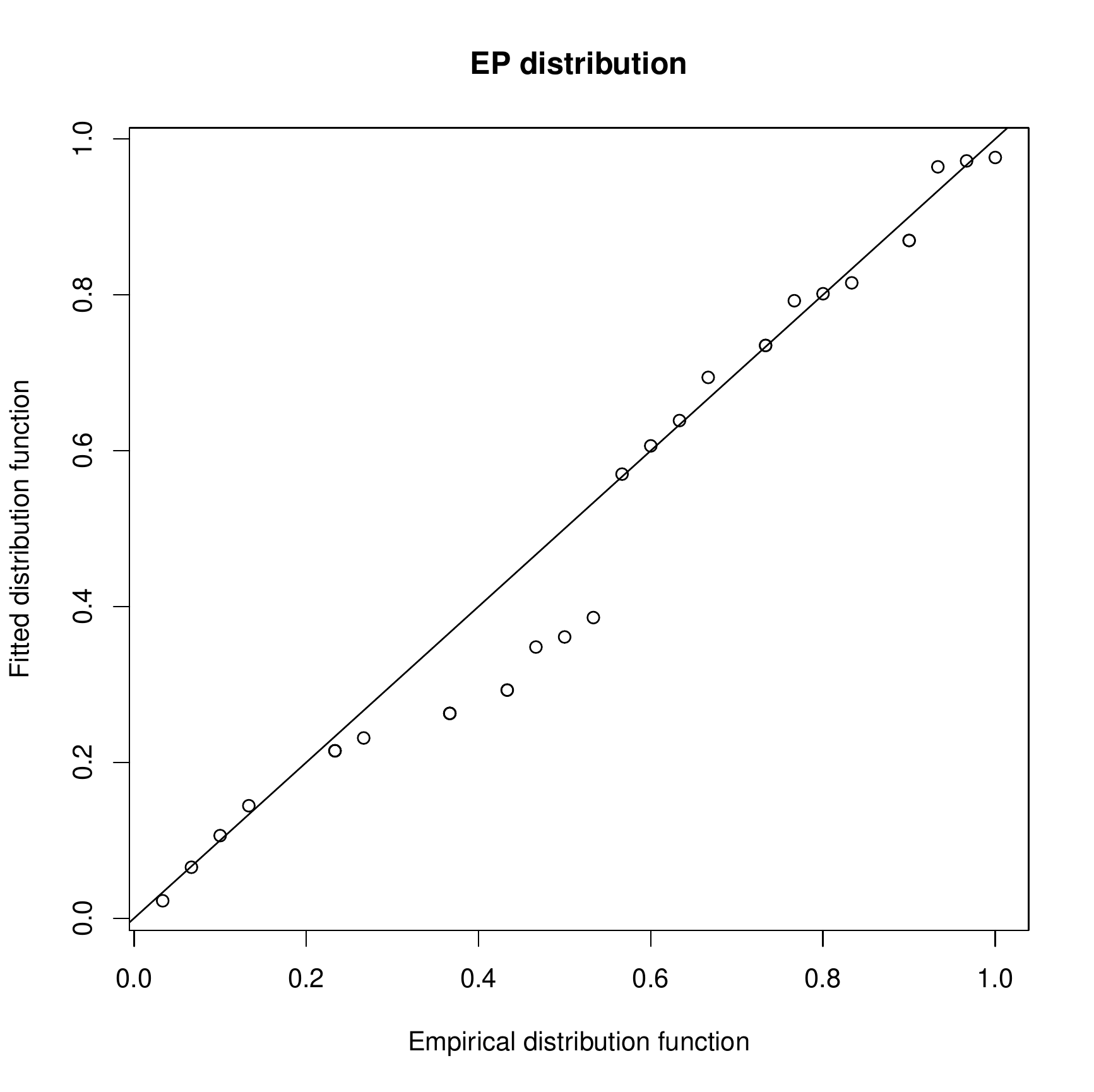}\includegraphics[width=0.5\textwidth]{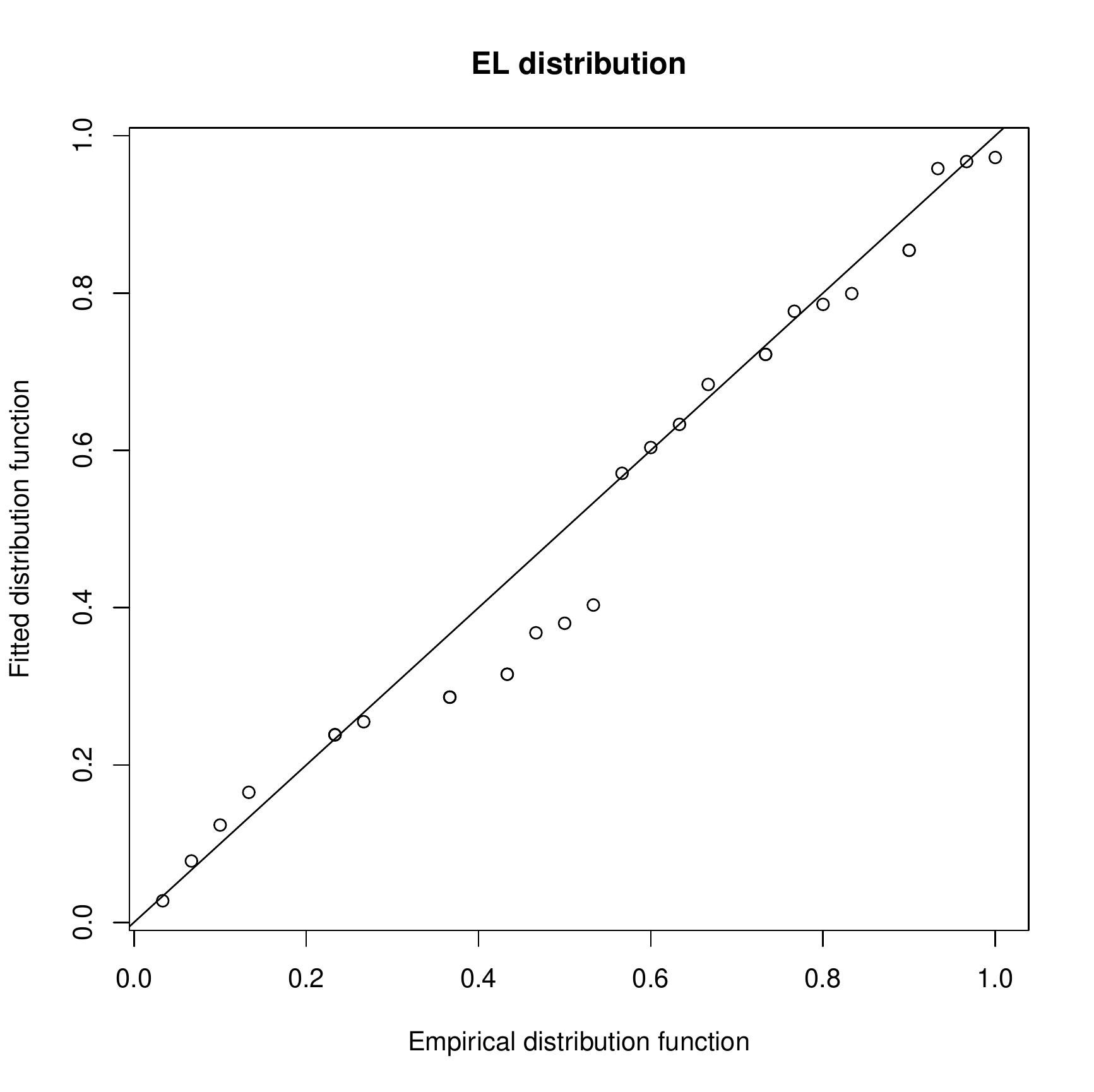}
		\includegraphics[width=0.5\textwidth]{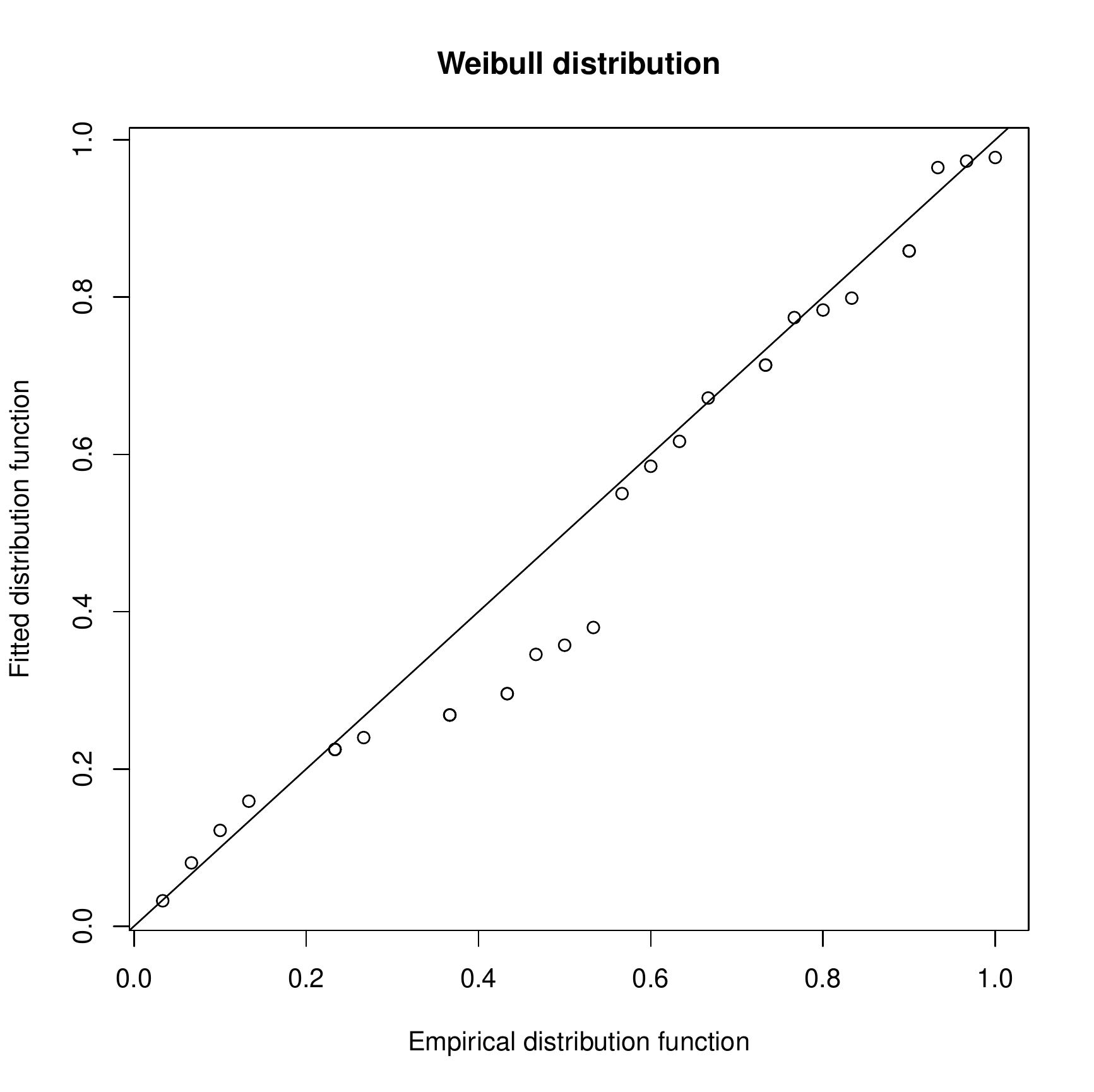}\includegraphics[width=0.5\textwidth]{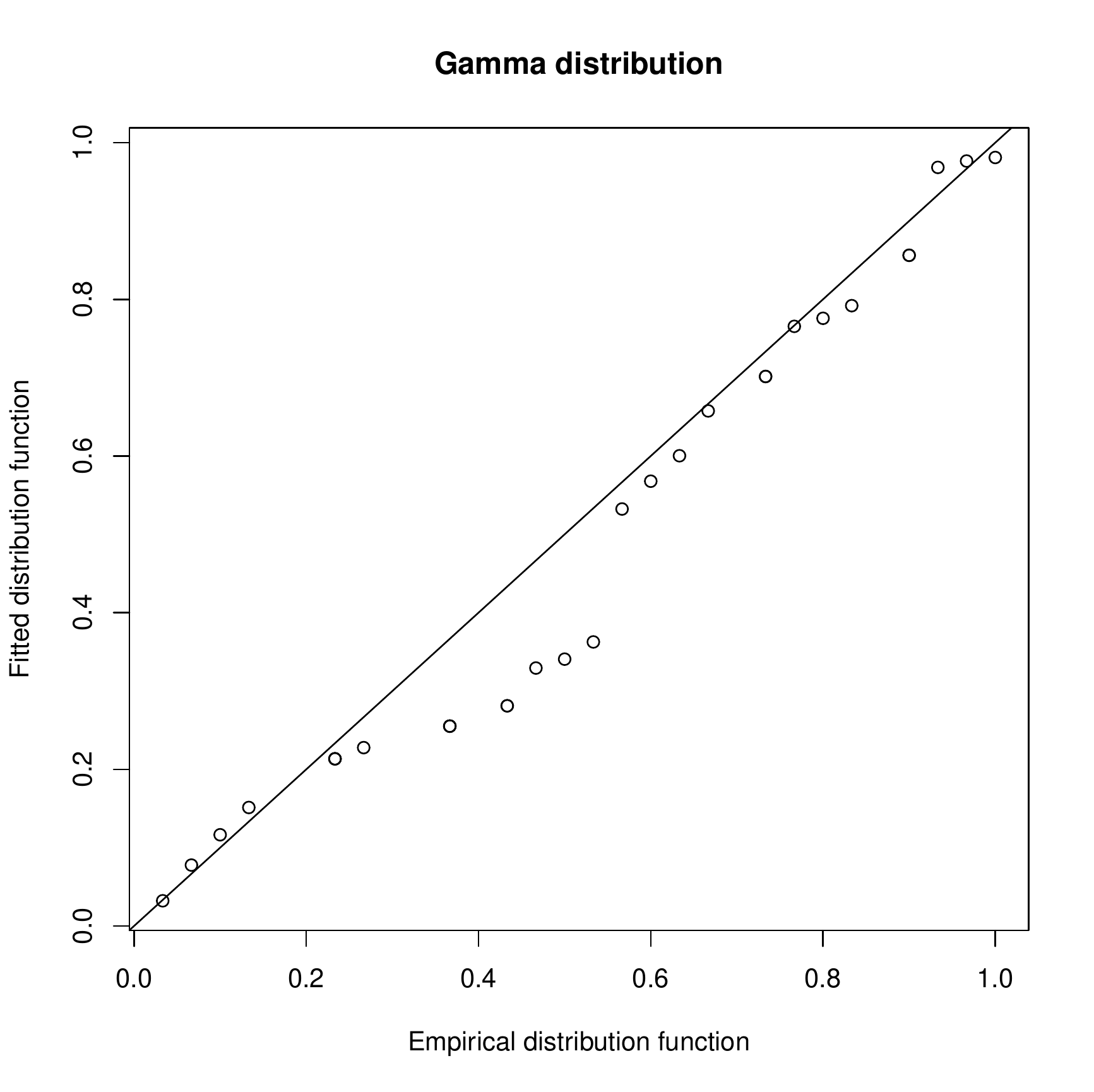}
	\caption{Empirical versus fitted distribution functions for the first data set.}
	\label{figureapll1}
\end{figure}

\begin{figure}[h!]
	\centering
		\includegraphics[width=0.5\textwidth]{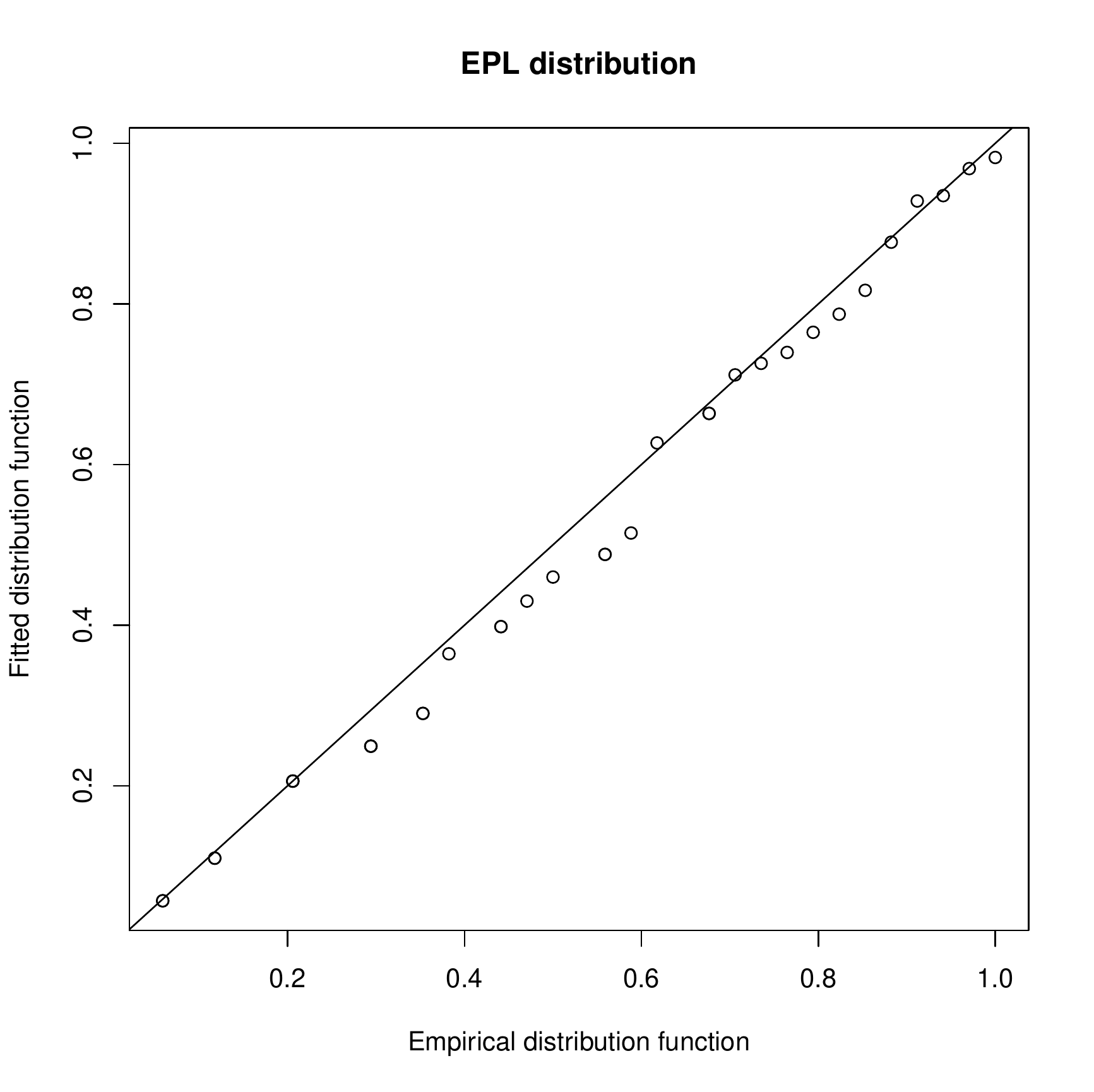}\includegraphics[width=0.5\textwidth]{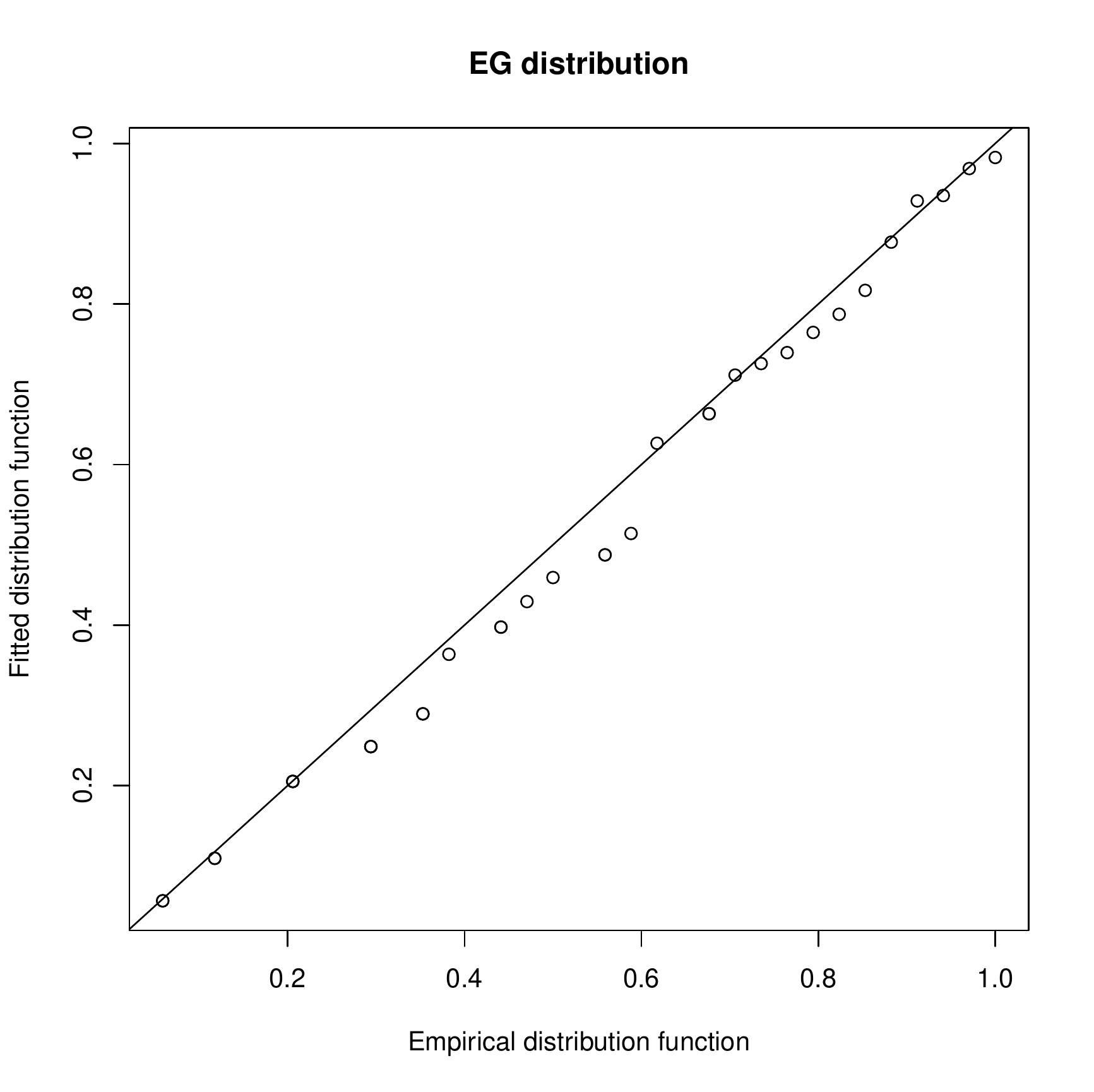}
		\includegraphics[width=0.5\textwidth]{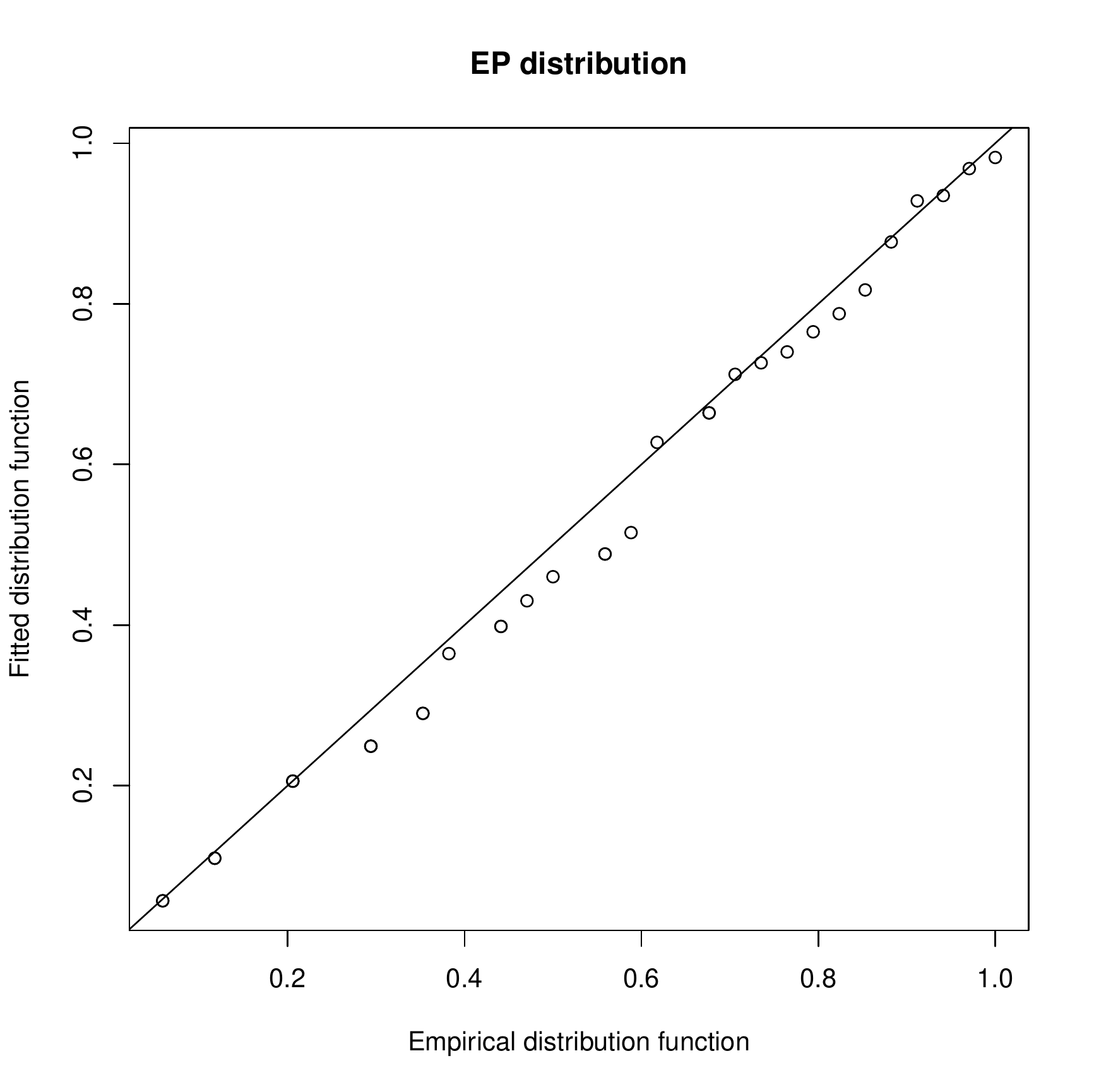}\includegraphics[width=0.5\textwidth]{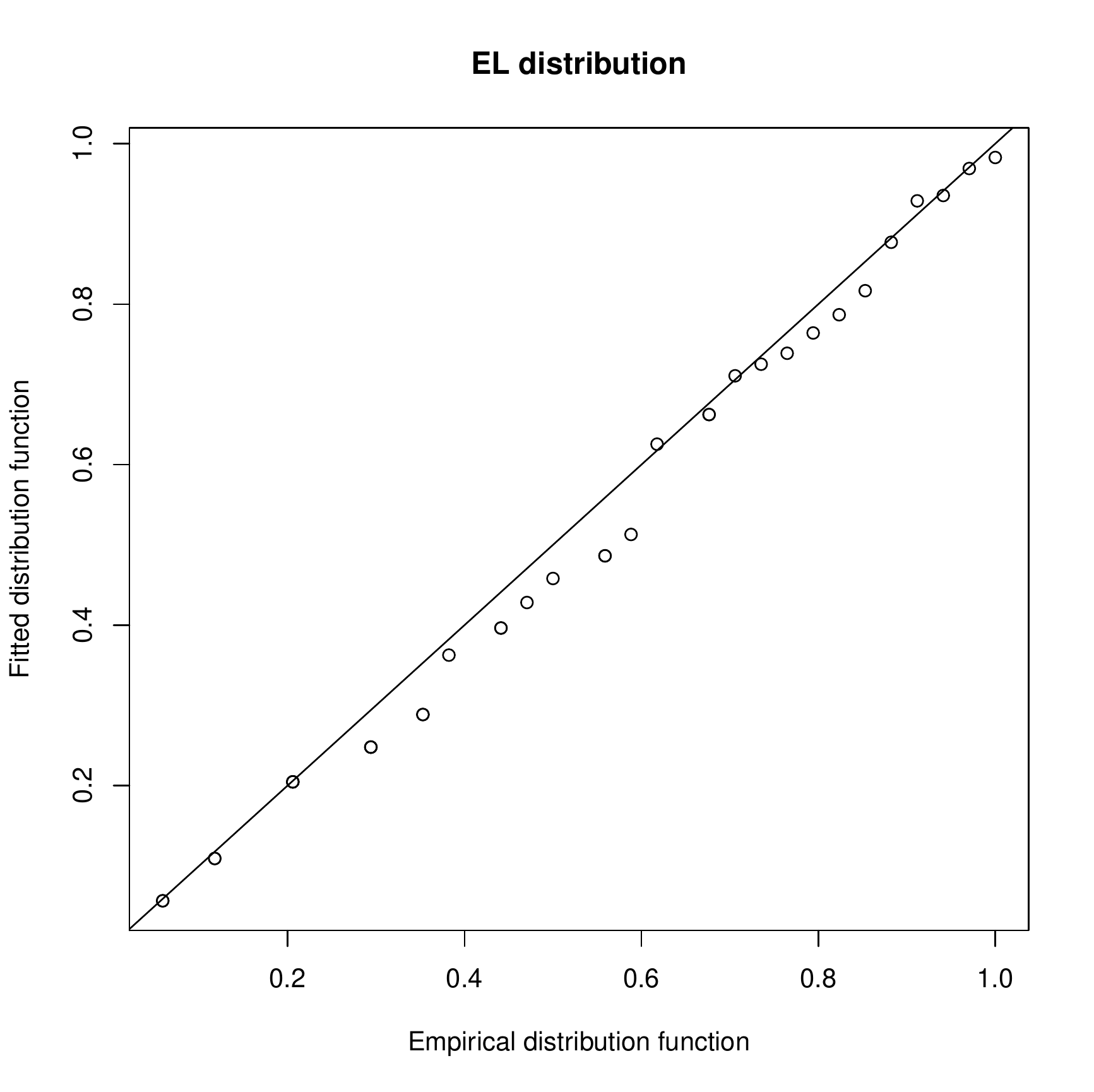}
		\includegraphics[width=0.5\textwidth]{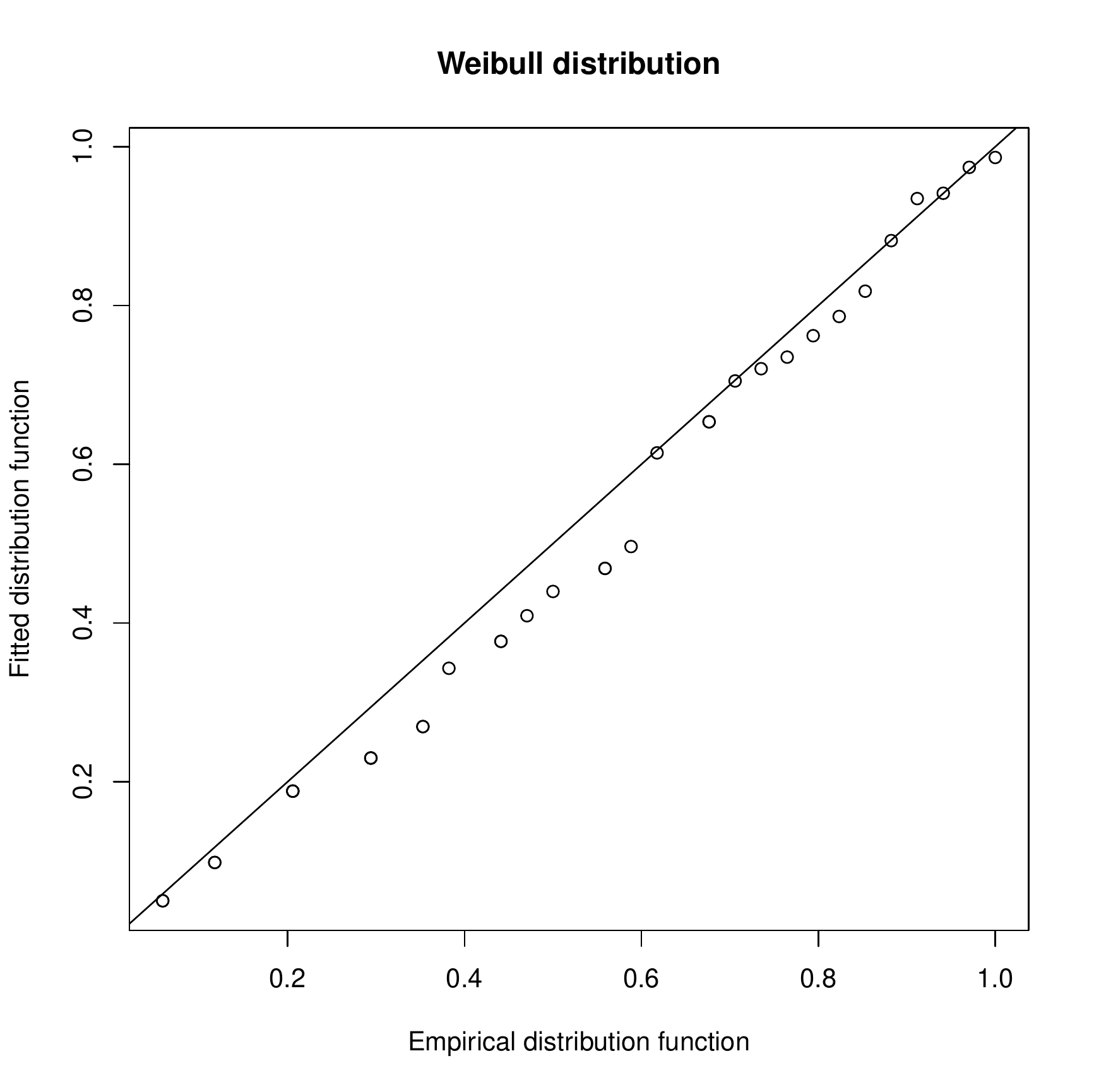}\includegraphics[width=0.5\textwidth]{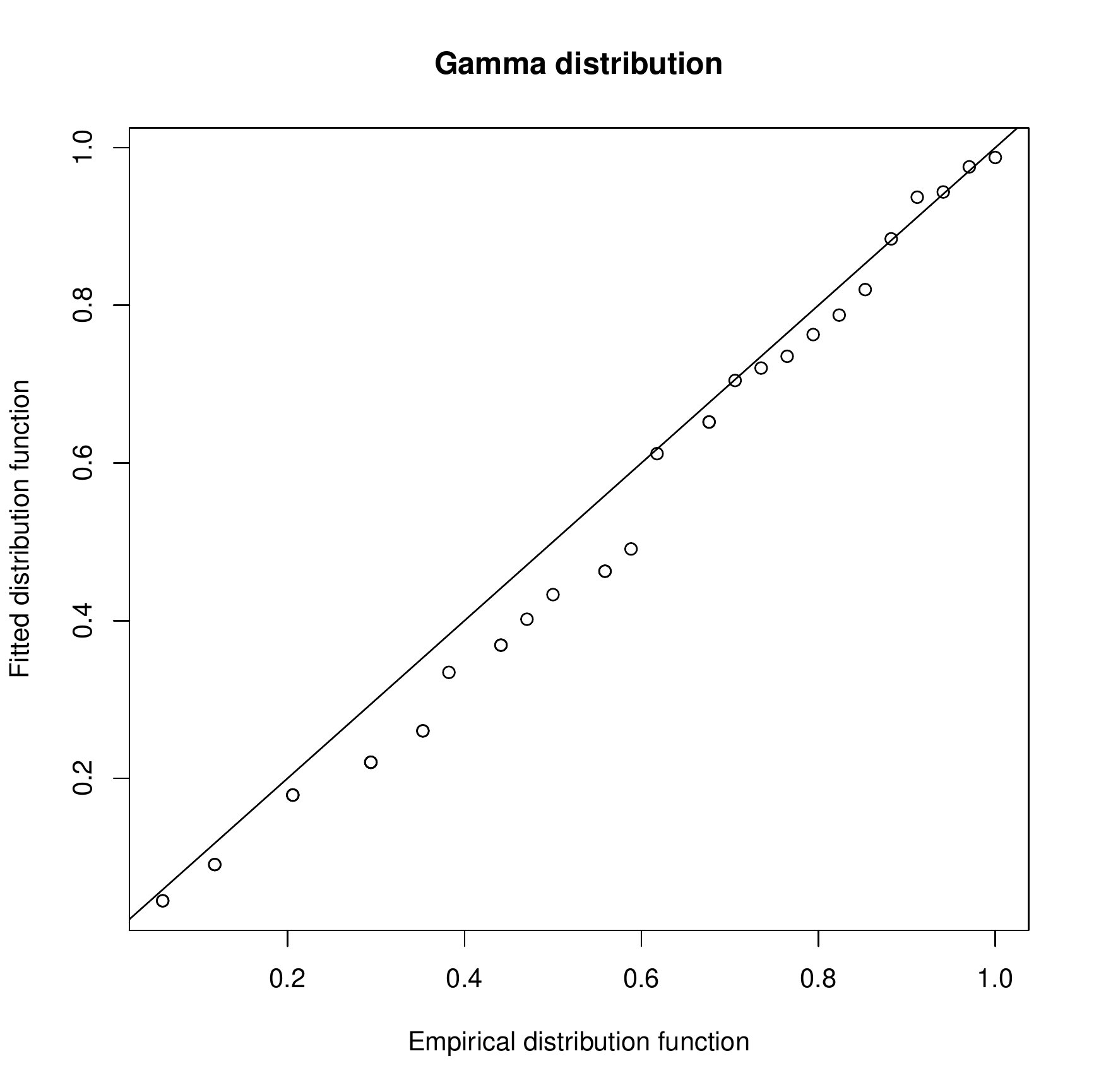}
	\caption{Empirical versus fitted distribution functions for the second data set.}
	\label{figureapll2}
\end{figure}

The good performance of our distribution is also confirmed by the Figure \ref{applfig1}. Figures \ref{figureapll1} and \ref{figureapll2} show empirical versus fitted distribution functions of the models considered for the two data sets. Once again, we see that EPL distribution presents a good fit as the other distributions considered.

\section{Conclusion}

In this paper, a new lifetime distribution with decreasing failure rate, named
exponential Poisson-Lindley (EPL) distribution, was introduced by compounding exponential and Poisson-Lindley distributions. Several statistical properties
of it were derived and discussed, in particular: mean lifetime, moments, order
statistics and R\'enyi entropy. Moreover, maximum likelihood estimation and Fisher's
information matrix were presented and discussed. Applications of the EPL distribution were carried out on two real
data sets. Based on the results of these applications, it is shown that the EPL distribution provides a good competitor among the other well-known lifetime distributions.

\end{document}